\documentclass{ifacconf}

\usepackage{amsmath,amssymb,amsfonts}
\usepackage{algorithmic}
\usepackage{graphicx}      
\usepackage{natbib}        
\usepackage{textcomp}
\usepackage{float}
\usepackage{breqn}
\usepackage{dblfloatfix} 
\usepackage{multicol}
\usepackage{mathtools}
\usepackage{color}

	\allowdisplaybreaks
\begin{document}
\begin{frontmatter}

\title{An MPC Approach to Transient Control of Liquid-Propellant Rocket Engines} 


\author[First,Third]{Sergio P\'{e}rez-Roca} 
\author[First]{Julien Marzat} 
\author[First]{\'{E}milien Flayac} 
\author[First]{H\'{e}l\`{e}ne Piet-Lahanier}
\author[Second]{Nicolas Langlois}
\author[Third]{Fran\c{c}ois Farago}
\author[Third]{Marco Galeotta}
\author[Fourth]{Serge Le Gonidec}

\address[First]{\textit{DTIS, ONERA, Universit\'{e} Paris-Saclay}, Chemin de la Huniere, 91120 Palaiseau, France (e-mails: \{sergio.perez\_roca, julien.marzat, emilien.flayac, helene.piet-lahanier\}@onera.fr)}
\address[Second]{\textit{Normandie Universit\'{e}, UNIROUEN,} \textit{ESIGELEC, IRSEEM}, Rouen, France (e-mail: nicolas.langlois@esigelec.fr)}
\address[Third]{\textit{CNES - Direction des Lanceurs}, 52 Rue Jacques Hillairet, 75612 Paris, France (e-mails: \{francois.farago, marco.galeotta\}@cnes.fr)}
\address[Fourth]{\textit{ArianeGroup SAS}, For\^{e}t de Vernon, 27208 Vernon, France, (e-mail: serge.le-gonidec@ariane.group)}

\begin{abstract}                

The current context of launchers reusability requires the improvement of control algorithms for their liquid-propellant rocket engines. Their transient phases are generally still performed in open loop. In this paper, it is aimed at enhancing the control performance and robustness during the fully continuous phase of the start-up transient of a generic gas-generator cycle. The main control goals concern end-state tracking in terms of combustion-chamber pressure and chambers mixture ratios, as well as the verification of a set of hard operational constraints. A controller based on a nonlinear preprocessor and on linear MPC (Model-Predictive Control) has been synthesised, making use of nonlinear state-space models of the engine. The former generates the full-state reference to be tracked while the latter achieves the aforementioned goals with sufficient accuracy and verifying constraints for the required pressure levels. Robustness considerations are included in the MPC algorithm via an epigraph formulation of the \textit{minimax} robust optimisation problem, where a finite set of perturbation scenarios is considered.

\end{abstract}


\end{frontmatter}

\section{Introduction}\label{intro}
In the current context of launcher vehicles design, reusability is considered as a major factor. From the automatic control perspective, the potential need for reusable liquid-propellant rocket engines (LPRE) implies stronger robustness requirements than controlling expendable ones due to their multi-restart and thrust-modulation capabilities. 
The classical multivariable control of main-stage LPRE had attained a reduced throttling envelope (70\%-120\%) in test benches. At real flights, only the nominal operating point is generally targeted. In the future European \textit{Prometheus} engine, it is aimed at throttling down to 30\% (\cite{baiocco_p._technology_2016}). Thus, an enlarged validity domain for reusability has to be conceived. At least, it becomes crucial to maintain tracking and robustness at those low throttle levels, where physical phenomena are more difficult to anticipate.\\
The main control problem in these multivariable systems primarily consists in tracking set-points in combustion-chamber pressure and mixture ratio, whose references stem from launcher needs. Control-valves opening angles are adjusted in order to adapt engine's operating point while respecting some constraints. The most common control approaches identified in the literature rely on linearised models about operating points for synthesising steady-state controllers, most of them based on PID techniques (such as \cite{nemeth_e._reusable_1991}). Generally, initial MIMO (Multi Input Multi Output) systems are considered decoupled into dominant SISO (Single Input Single Output) subsystems. 
Off-line optimisation strategies have also been carried out (\cite{dai_x._damage-mitigating_1996}). More complex approaches present in the literature, incorporating some nonlinear (\cite{lorenzo_c.f._nonlinear_2001}), hybrid (\cite{musgrave_j.l._real-time_1996}) or robust (\cite{saudemont_r._study_2000}) techniques, enhance certain aspects of performance and robustness. And in the event that a component fails, some reconfiguration-control strategies have been proposed (\cite{musgrave_j.l._real-time_1996}).\\
To the best of our knowledge, there are no publications which consider not only the steady state but also the demanding transient phases at the same level of performance and robustness, as reviewed in \cite{perez-roca_s._survey_2019}. 
Pre-defined sequences of engine operation (start-up and shutdown), are traditionally managed in open loop with narrow correction margins. They consist in an initial succession of discrete events including valves openings and chambers ignitions. Once these commands have all been executed, the second part of the transient, which is completely continuous, takes place until the steady state is achieved. The main reasons for performing open-loop (OL) control in the first (discrete-event) phase, explained in \cite{nemeth_e._reusable_1991}, are controllability and observability issues at very low mass flows. Transient control through valves starts to be plausible once all events have finished. This observation has been considered in this paper, where only the second part of the start-up transient, fully continuous, is controlled.\\
The objective of this work is to control the start-up transient of a pump-fed LPRE. In this case, a gas-generator-cycle engine is targeted. Concretely, it is aimed at achieving combustion-pressure and mixture-ratio end-state tracking while complying with hard operational constraints, mainly concerning mixture ratios, turbopumps rotational speeds and valves actuators angular velocities. The control strategy presented in this paper is based on Model Predictive Control (MPC), which is accompanied by a preprocessor for full-state reference generation. This method was selected as the most appropriate for this kind of complex systems with hard operational constraints, as introduced in the next sections. Indeed, it is more and more used in industry and can be extended for instance with robustness (\cite{mayne_d.q._constrained_2000}) or hybrid considerations, which will be interesting for future work on this topic.\\
This paper is organised as follows. Section \ref{model} serves as a recapitulation of modelling considerations published in a previous paper. The state-space system used in the following sections is presented there. Section \ref{ctrl} describes the control strategy carried out, which mainly consists in the use of MPC techniques. Section \ref{resu} depicts results and includes their analysis. Finally Section \ref{conclu} serves as a conclusion.

\section{Modelling}\label{model}
\begin{figure}[h]
	\begin{center}
		\includegraphics[width=0.4\textwidth]{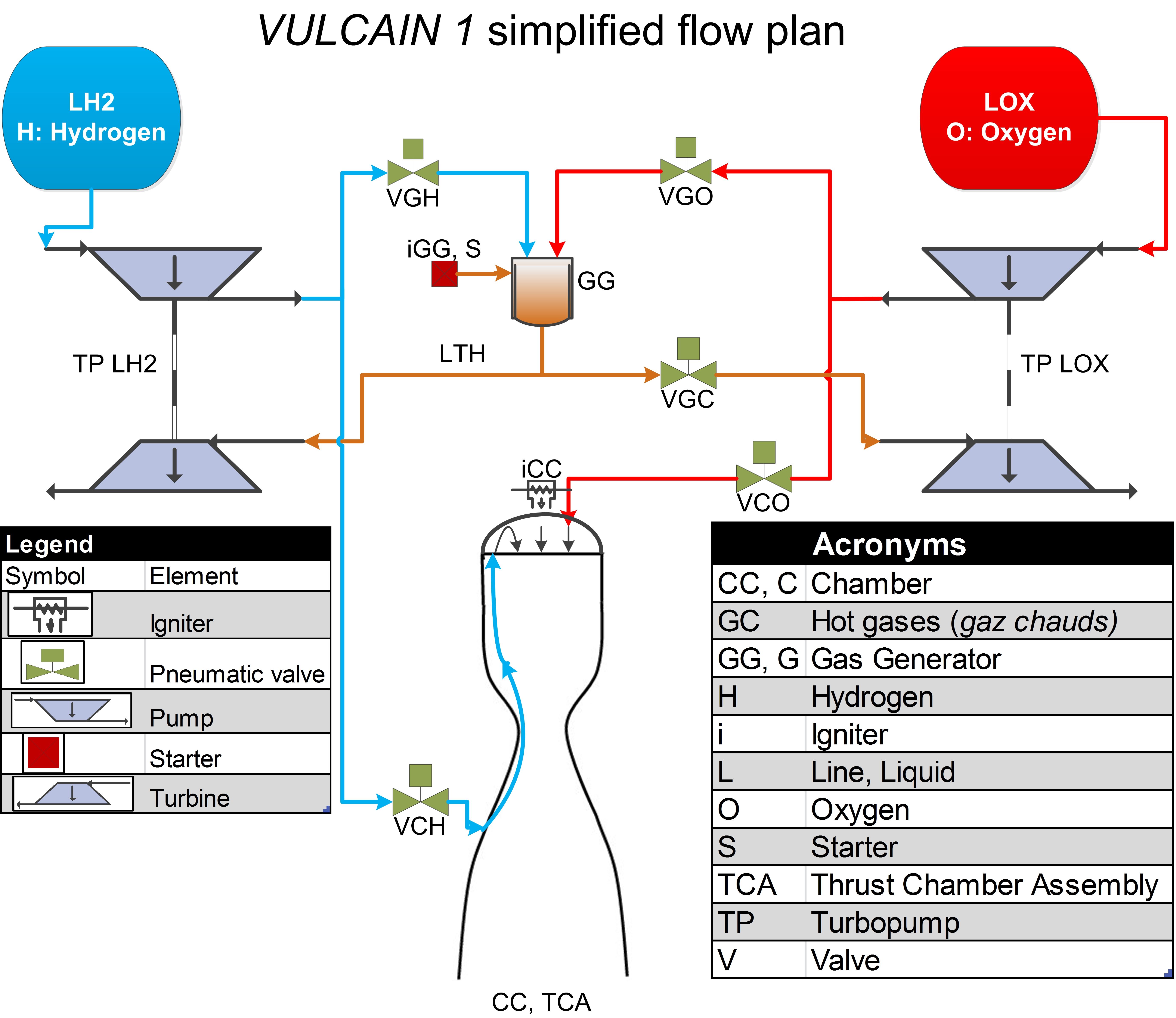}
		\caption{\textit{Vulcain 1} simplified flow plan}
		\label{figmodel}
	\end{center}
\end{figure}

The modelling strategy used in this paper was introduced in \cite{perez-roca_s._derivation_2018}. Several types of models are employed in the control loop in this paper. Concerning the plant on which the control is exerted, a simulator of the real plant was constructed in first place. 
This simulator, whose structure is built component-wise, already considers the basic thermo-fluid-dynamics and mechanics of LPRE elements: mass, energy and momentum conservation equations. The engine considered in this paper, representative of the \textit{Vulcain 1}, presents a gas-generator (GG) cycle. In Fig. \ref{figmodel}, its main components are depicted and the main acronyms are summarised. It consists in a $LOX/LH_2$ (liquid oxygen as oxidiser, liquid hydrogen as fuel) engine. The hot-gas flow necessary to drive turbines comes from a GG, a small combustion chamber that receives a small portion of the main propellant flow. The actuators considered in this paper are five continuously-controllable valves (VCH, VCO, VGH, VGO and VGC). Apart from those, there are two discrete actuators: one binary igniter ($i_{CC}$) and one binary starter ($i_{GG}$). However, they are considered as active in this paper in order to treat the continuous part ot transients (up from $1.5s$ after start command), where only continuous control takes place. The GG starter injects hot gas into that cavity during less than $1.5s$ so as to start driving turbines.
Valves angles ($\alpha$), which have a nonlinear but direct relation to sections ($A$), control the flows to the main combustion chamber (VCH and VCO), 
to the GG (VGH, VGO), and to the oxidiser turbine (VGC). The latter consists in the main contribution in determining mixture ratio ($MR$), which is defined as the quotient between oxidiser and fuel mass flow rates $MR = \dot{m}_{ox}/\dot{m}_{fu}$. This ratio, a major performance indicator in LPRE, is established at three levels: at an engine's global level ($MR_{PI}$), taking pumped propellants into account; in the combustion chamber ($MR_{CC}$) and in the GG ($MR_{GG}$).\\
The simulator was then translated into a nonlinear state-space model by joining components equations symbolically.
At this stage, having already carried out certain simplifications with respect to the initial simulator, the model is referred to as \textit{complex NLSS} (nonlinear state-space) or $f_c(\mathbf{x},\mathbf{u})$. However, this model was too cumbersome for control design.
Hence, it was further reduced until achieving the here-called \textit{simplified NLSS} such that $\dot{\mathbf{x}}=f_s(\mathbf{x},\mathbf{u})$, more tractable for its manipulation and derivation of control laws.
The cost of these simplifications, such as the consideration of constant thermodynamic properties, is the increase of modelling error. A difference in comparison with \cite{perez-roca_s._derivation_2018} is the consideration of cavity temperatures as correlated functions of the respective $MR$, thereby shortening the state vector. Besides, in this paper, all equations, states and control have been rendered non-dimensional with respect to the nominal equilibrium values. In terms of notation, the presence of a tilde ($\tilde{}$) on top of a quantity means that it is dimensional and its absence means the contrary.\\
The number of states is $n=12$ and $m=5$ is the number of control inputs. Here, the state vector $\mathbf{x}$, of both NLSS, comprises the two turbopumps speeds $\omega_H$ and $\omega_O$, the four pressures in the system ($p_{CC}$ of combustion chamber, $p_{GG}$ of the GG, $p_{LTH}$ for hydrogen-turbine inlet cavity and $p_{VGC}$ for oxygen-turbine inlet cavity) and six mass flows, including the ones streaming through control valves ($\dot{m}_{VCH}$, $\dot{m}_{VCO}$, $\dot{m}_{VGH}$, $\dot{m}_{VGO}$ and $\dot{m}_{VGC}$) and the one streaming through the hydrogen-turbine inlet pipe $\dot{m}_{LTH}$.
\begin{multline}
\mathbf{x}=[\begin{matrix}
\omega_H & \omega_O & p_{CC} & p_{GG} & p_{LTH} & p_{VGC} &  \dot{m}_{LTH} & \dot{m}_{VCH}   \end{matrix}\\
\begin{matrix}\dot{m}_{VCO} & \dot{m}_{VGH} & \dot{m}_{VGO} &	\dot{m}_{VGC}
\end{matrix}]^T.
\end{multline}
The states with greater tracking relevance are incorporated into a reduced state vector $\mathbf{x}_z$:
\begin{equation}
\mathbf{x}_z=[\begin{matrix} p_{CC} & \dot{m}_{VCH} & \dot{m}_{VCO} & \dot{m}_{VGH} & \dot{m}_{VGO} \end{matrix}]^T.
\end{equation}
The control input $\mathbf{u}$ contains the sections of the five control valves:
\begin{equation}
\mathbf{u}=[\begin{matrix} A_{VCH} & A_{VCO} & A_{VGH} & A_{VGO} & A_{VGC} \end{matrix}]^T.
\end{equation}
Modelling error is specially present in mass flows, which can present a mismatch of 10 to 25\% at each step of simplification (simulator, $f_c$, $f_s$ and linearised models). Errors in the other states are generally below 10\% at each step. The state is assumed to be fully measurable in the real engine. This is a realistic assumption for $\omega$ and $p$. However, measuring some mass flows would be problematic in terms of engine design. Mass flows are generally not measured in LPRE, but estimated through pressure, temperature and volumetric flow measurements. This estimation process is deemed perfect in this paper.

\section{Controller design}\label{ctrl}
\begin{figure*}[h]
	\begin{center}
		\includegraphics[width=0.65\textwidth]{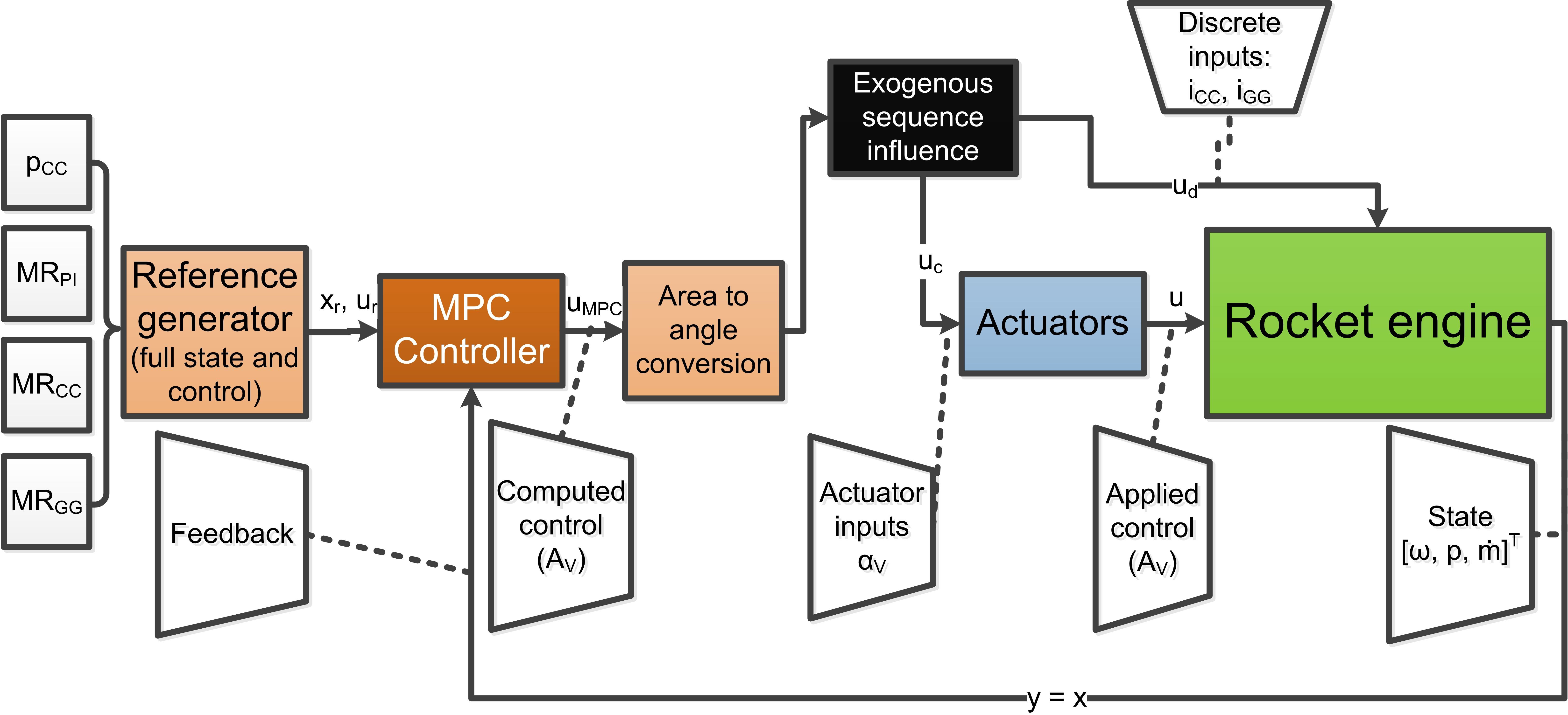}
		\caption{Control-loop diagram}
		\label{ctrl_loop}
	\end{center}
\end{figure*}

The goal of the controller is to drive the state towards a desired reference $\mathbf{x}_{r}$ at the end of the start-up transient, with a special focus on having a small tracking error in $\mathbf{x}_z$. At the same time, a set of hard constraints on $\mathbf{x}$ and $\mathbf{u}$ has to be met throughout the transient. This second goal is somewhat more important than the former in order to avoid excessive temperatures, $p$ or $\omega$ during engine's operation. The duration of the start-up transient until reaching the reference is required to range between $2$ and $4s$, which allows the system to cope with possible perturbations or uncertainties while complying with constraints. A concrete reference trajectory is not imposed.\\
The designed controller comprises two main parts. The main component is the MPC block, which receives a full-state reference and drives the system to it while satisfying constraints. The other part of the controller consists in a preprocessing block, which serves to generate the full reference vector $\mathbf{x}_{r}$ fed to the MPC. The whole control diagram is depicted in Fig.\ref{ctrl_loop}. The remaining elements in that diagram are the following. To the right there is the simulation of the rocket engine (complex simulator), run at $10^{-5}s$ for capturing fast dynamics and for being robust to numerical stiffness. The inputs of that simulator and of the state-space model used for control are valve sections $\mathbf{u}$. However, the actuators model (internal valve actuators) requires an input in terms of $\alpha$. That is the reason why there is a conversion block, characterised by static and monotone nonlinear functions. The MPC controller provides valve sections that are then translated into angles. The cause for considering valve actuators as a separate entity is the fact that they represent an internal servo-loop, in which the angular position of the valve is tuned by means of a hydraulic or electrical actuator, modelled as a second-order system. 

\subsection{Preprocessing}\label{preproc}

The preprocessing block serves as an off-line reference generator for the MPC controller. This is required because the set of reference commands derived from launcher needs is not sufficient to provide a complete-state target equilibrium point to the engine controller. Hence, a way of restoring a full state vector from those data is necessary. In addition, without this $\mathbf{x}_r$, the posterior MPC controller would not attain the tracking goal with high precision. This is mainly due to the fact that $f_s$ is linearised about $\mathbf{x}_r$, as explained in Section \ref{mpc}.\\
As said before, $\mathbf{x}$ presents twelve states, and there are four reference inputs: $p_{CC,r}$, $MR_{PI,r}$, $MR_{CC,r}$ and $MR_{GG,r}$. Moreover, the last three, in contrast to $p_{CC}$, do not directly correspond to states in the model. They establish relations between $\tilde{\dot{m}}_i$. In a first place, thanks to the selected pressure and $MR_{CC,r}$, the choked-flow static equation
\begin{equation}
	\dot{m}_{CC}=\frac{p_{CC} A_{th}}{C^*}
\end{equation}
can provide $\dot{m}_{VCH}$ and $\dot{m}_{VCO}$, where $A_{th}$ is the throat area and $C^*$ is the characteristic velocity, dependent on $MR$ itself. Then, the rest of states at equilibrium are computed by solving the following overdetermined system of nonlinear equations:
\begin{equation}
\begin{cases}
	\dot{\mathbf{x}}=f_c(\mathbf{x}_r,\mathbf{u}_r)=0 \quad \setminus (\dot{p}_{CC}=0)\\
	\frac{\tilde{\dot{m}}_{VCO}+\tilde{\dot{m}}_{VGO}}{\tilde{\dot{m}}_{VCH}+\tilde{\dot{m}}_{VGH}}=MR_{PI,r}\\
	\frac{\tilde{\dot{m}}_{VGO}}{\tilde{\dot{m}}_{VGH}}=MR_{GG,r}\\
	\tilde{\dot{m}}_{VGH}+\tilde{\dot{m}}_{VGO}=\tilde{\dot{m}}_{LTH}+\tilde{\dot{m}}_{VGC}.
\end{cases}
\end{equation}
The first equations force the ODE to be at equilibrium, the second and the third ones determine the $MR$ and the last one enforces the equilibrium of $\sum \tilde{\dot{m}}_i$ in the GG. The ODE for $\dot{p}_{CC}$ is removed since it is completely dependent on the reference inputs, not providing additional information. This resolution is performed numerically via nonlinear least squares due to the unavailability of an analytic solution of the system, of either $f_c$ or $f_s$. The complex model has been chosen to increase accuracy.

\subsection{MPC algorithm}\label{mpc}

MPC predicts the future system behaviour along a horizon, and optimises control inputs according to a cost function generally related to a reference trajectory or to an end state. In this paper, the dynamic model used in the state-feedback MPC controller is considered as a linearisation of $f_s$ about the previously computed $(\mathbf{x}_r,\mathbf{u}_r)$ and as a zero-order hold discretisation at $\Delta t=10ms$ (due to computational constraints):
\begin{equation}
	\mathbf{\Delta x}_{k+1} = A_d(\mathbf{x}_r,\mathbf{u}_r)\mathbf{\Delta x}_k + B_d(\mathbf{x}_r,\mathbf{u}_r)\mathbf{\Delta u}_k
\end{equation}
Thus, in linear terms, the goal of the controller is to find the set of $\mathbf{\Delta u}=\mathbf{u}-\mathbf{u}_r$ that drives the state to $\mathbf{\Delta x}=\mathbf{x}-\mathbf{x}_r=\mathbf{0}$. The matrix $A_d$ is stable for all the physically feasible $\mathbf{x}_r$, which is a particularity of GG-cycle LPRE. In this MPC section, in order to lighten notation, all $\mathbf{x}$ and $\mathbf{u}$ refer to variations with respect to the equilibrium point. The approach carried out is partially based on the quasi-infinite horizon (QIH) approach by \cite{chen_h._quasi-infinite_1998}, because it presents proofs for guaranteed stability and end-state reachability of MPC by incorporating the notion of a terminal region. The MPC drives the system to that region, where a fictitious local controller $K$ performs the precise tracking at the end of the state prediction horizon, $N_p+1$. 
However, in MPC only the first computed control, $\mathbf{u}_{MPC}\equiv\mathbf{u}_{1}$, is transmitted to the plant. Hence, the real role of the fictitious feedback $\mathbf{u}_{N_p+1}=K\mathbf{x}_{N_p+1}$ is to compute the $P$ matrix of a Lyapunov function $V(\mathbf{x})=\mathbf{x}^T P\mathbf{x}$, by solving the following Lyapunov equation:
\begin{equation}\label{lyap}
	(A_K+\kappa I)^TP+P(A_K+\kappa I)=-Q_K-K^T R_K K.
\end{equation}
In (\ref{lyap}), the compound of the linear system with a simple LQR feedback controller is considered, $A_K=A_c+B_cK$ (where $A_c$ is the continuous counterpart of $A_d$), $\kappa\in\mathbb{R}^+$ (satisfying $\kappa<-\lambda_{max}(A_K)$) and $Q_K$ and $R_K$ are positive definite symmetric matrices $Q_K\in\mathbb{R}^{n\times n}$, $R_K\in\mathbb{R}^{m\times m}$. The computed $P\in\mathbb{R}^{n\times n}$ serves to add an additional terminal-region term in the MPC cost. In addition, an integral action is also included to enforce a more precise tracking on $\mathbf{x}_z$. Those integral decision variables are denoted by $\mathbf{z}$ and present a corresponding weight matrix $S\in\mathbb{R}^{n_z\times n_z}$ in the cost, whose diagonal is $\left[1, 0.1, 0.1, 0.1, 0.1\right]$. Thus, the MPC cost $J$ is defined as:
\begin{dmath}
	\setlength{\abovedisplayskip}{1pt}
	\setlength{\belowdisplayskip}{1pt}
	J(\mathbf{x},\mathbf{u},\mathbf{z}) = \left(\sum_{k=1}^{N_p}\mathbf{x}_{k}^T Q\mathbf{x}_{k} + \sum_{k=1}^{N_u}\mathbf{u}_k^T R\mathbf{u}_k+\sum_{k=1}^{N_p}\mathbf{z}_k^T S \mathbf{z}_k\right)\Delta t + \mathbf{x}_{N_p+1}^T P\mathbf{x}_{N_p+1} ,	
\end{dmath}
which consists in the traditional quadratic cost on states and controls plus the integral and terminal costs, with a prediction horizon $N_p=10$ steps ($0.1s$) and a control horizon $N_u=5$. Implicitly, the last control $\mathbf{u}_{N_u}$ is used for $k\geq N_u$. Further extensions of these horizons did not improve the solutions in terms of tracking or constraints satisfaction. $Q$ and $R$ are positive-definite symmetric weighting matrices $Q\in\mathbb{R}^{n\times n}$, $R\in\mathbb{R}^{m\times m}$, whose diagonals have been computed off-line via Kriging-based black-box optimisation as in \cite{marzat_j._automatic_2010}. The criterion for that weight selection concerns the minimisation of static error and overshoot in simulations.\\
Furthermore, the first steps towards a robust consideration of the problem have been implemented. The minimisation of the previous $J$ under constraints is not robust. Indeed, robustness to parameters and initial conditions variations, perturbations and modelling error is very important in this application. Robust MPC approaches generally make use of the \textit{minimax} optimisation, which minimises the worst-case scenario. A generic expression of this problem is the following, in which $\mathbf{w}$ represents disturbance (\cite{mayne_d.q._constrained_2000}):
\begin{equation}\label{minimax}
\setlength{\abovedisplayskip}{2pt}
\setlength{\belowdisplayskip}{2pt}
	\begin{aligned}
	\min_{\mathbf{u}}\max_{\mathbf{w}} \quad & J(\mathbf{x},\mathbf{u})\\
	\textrm{s.t.} \quad & \mathbf{x}\in X \quad \forall \mathbf{w}\in\mathbb{W}^n \\
	& \mathbf{u}\in U \quad \forall \mathbf{w}\in\mathbb{W}^n\\
	\end{aligned}
\end{equation}
However, solving (\ref{minimax}) for all possible perturbations is too computationally costly for this application. Hence, it has been opted for choosing a finite set of disturbance scenarios (in a similar manner to \cite{calafiore_g.c._robust_2013}) and for solving an equivalent formulation based on \cite{loefberg_j._minimax_2003}. Concretely, it consists in minimising $\gamma\in\mathbb{R}^+$ via an epigraph formulation. In this paper, that $\gamma$ constrains the $J$ of the original problem evaluated at several perturbed states propagations $\mathbf{x}_{i}$:
\begin{equation}\label{xi}
\setlength{\abovedisplayskip}{2pt}
\setlength{\belowdisplayskip}{2pt}
	\begin{aligned}
	\mathbf{x}_{i} = [\mathbf{x}_{i,1},...,\mathbf{x}_{i,k},...,\mathbf{x}_{i,N_p+1}]^T,\quad i\in I\\
 	\mathbf{x}_{i,k+1}=A_d\mathbf{x}_{i,k}+B_d \mathbf{u}_k+\mathbf{w}_{i,k},\quad k\in[0,N_p+1],
 	\end{aligned}
\end{equation}
where $\mathbf{w}_{i,k}$ are certain selected perturbation vectors belonging to $\mathbb{W}=\{\mathbf{w}_{i,k},i\in I, k\in[0,N_p+1]\}$. $I$ is a finite set, which serves to index the considered perturbation cases. Indeed, the epigraph formulation allows to entirely shift the robustness considerations into the list of constraints. Therefore, only a smooth convex nonlinear programme (NLP) is required, which is more computationally tractable than (\ref{minimax}).
The minimisation problem proposed here, in which decision variables are extended to consider all $\mathbf{x}_{i}$, is described below:
\begin{align}\label{algo}
\min_{\mathbf{x}_{i},\mathbf{u},\mathbf{z}_i,\gamma} \quad & \gamma\\
\textrm{s.t.} \quad & J(\mathbf{x}_i,\mathbf{u},\mathbf{z}_i)\leq\gamma \quad \forall i\in I\notag\\
& \mathbf{x}_i\in X, \quad \mathbf{u}\in U \quad \forall i\in I\notag\\
&A_{ineq}[\mathbf{x}_i \quad \mathbf{u}]^T\leq \mathbf{b}_{ineq}  \quad \forall i\in I \notag\\
&A_{eq}[\mathbf{x}_i \quad \mathbf{u}]^T= \mathbf{b}_{i,eq} \quad \forall i\in I\notag\\
&\mathbf{x}_{i,N_p+1}^T P\mathbf{x}_{i,N_p+1}\leq \alpha_P \quad \forall i\in I    \notag\\
&\mathbf{z}_{i,k+1} = \mathbf{z}_{i,k} + \Delta t K_I\mathbf{x}_{z,i,k} \quad \forall i\in I, k\in[0,N_p].\notag \notag
\end{align}
\begin{figure}[b]
	\begin{center}
		\includegraphics[width=0.36\textwidth]{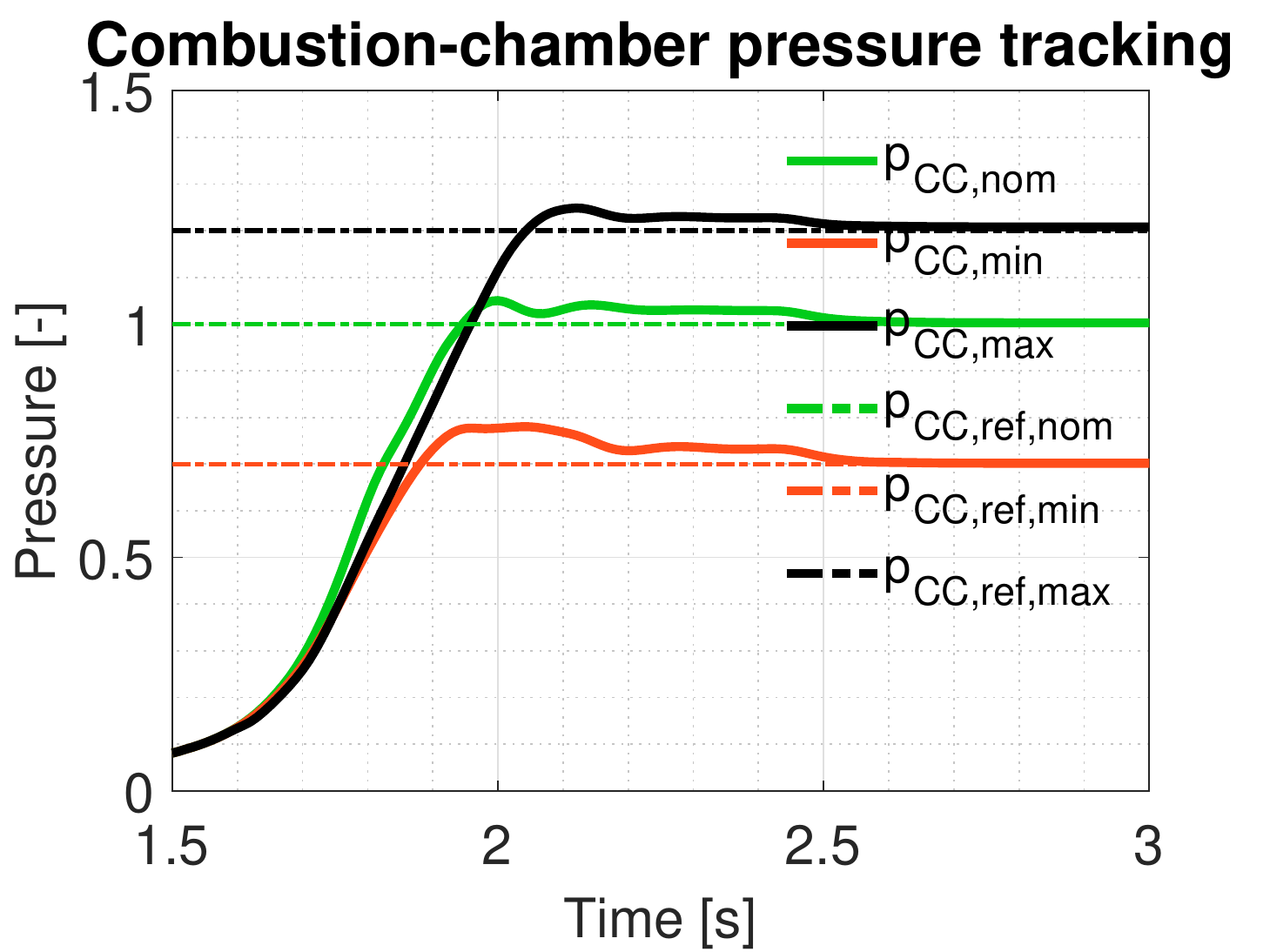}
		\caption{Tracking results in $p_{CC}$ for $p_{CC,r}=1$ (nominal), $p_{CC,r}=0.7$ (minimum) and $p_{CC,r}=1.2$ (maximum)}
		\label{resu_pCC}
	\end{center}
\end{figure}
$X$ and $U$ are the allowable sets for states and control (compact subsets of $\mathbb{R}^{n(N_p+1)}$ and $\mathbb{R}^{mN_u}$ respectively). The set $U$ for the first control $\mathbf{u}_{MPC}$ is specially constrained to comply with actuators capacity (\cite{luo_y._model_2004}):
\begin{equation*}
	\mathbf{u}_{MPC}\in [\max(\underline{U},\mathbf{u}_0-\dot{\mathbf{u}}_{max}\Delta t),\min(\overline{U},\mathbf{u}_0+\dot{\mathbf{u}}_{max}\Delta t)],
\end{equation*}
where $\mathbf{u}_0$ is the previous-step control (warm start is performed) and $\dot{\mathbf{u}}_{max}$ is the maximum sectional velocity of valves. Regarding the rest of constraints, (\ref{algo}) contains equality constraints (defined by $A_{eq}$ and $\mathbf{b}_{i,eq}$) for linear dynamics (\ref{xi}) and also linear inequality constraints (defined by $A_{ineq}$ and $\mathbf{b}_{ineq}$), for complying with $MR$ and actuators sectional-velocity bounds at all $\mathbf{x}_{i}$. In relation to the terminal region, a constant $\alpha_P$ refers to the neighbourhood in which the Lyapunov term of $J$ is constrained in a nonlinear way (further details on the QIH method in \cite{chen_h._quasi-infinite_1998}). The different $\mathbf{w}_{i,k}$ represent the various ways in which the system can evolve after an unknown perturbation or uncertainty in the state, and hence it is proposed to estimate them by analysing the modes of the system (eigenvectors of $A_c$). The total number of perturbation cases $I=\{1,2,3\}$ corresponds to a subset of the eigenvectors.
In this manner, the structural information of $A_c$ is used to define 
unfavourable disturbance scenarios, similarly to \cite{yedavalli_r.k._improved_1985}. The modulus of the vectors $\mathbf{w}_{i,k}$ is kept equal to $0.1$. It is important to emphasise the fact that the resulting $\mathbf{u}$ obtained in (\ref{algo}) has been confronted to all these perturbation scenarios and that all propagated perturbed states must comply with all constraints, thereby improving the robustness of the controller. This approach with equality constraints within an uncertain problem is only valid because of the finite choice of $\mathbf{w}_{i,k}$. The last line in constraints corresponds to the integrator dynamics (\cite{santos_l.o._-line_2001}), where $K_I$ is a gain matrix computed off-line in the same manner as $Q$ and $R$.

\section{Analysis of results}\label{resu}

The interior-point optimisation software \textit{IPOPT} (\cite{waechter_a._implementation_2006}) has been used to solve this smooth convex NLP within the \textit{MATLAB} environment. Simulations of the previously presented control loop are run from $1.5s$ until $3s$ after the start command, that is to say, during the time window in which continuous control is possible in engine start-up transient.
\begin{figure}[b]
	\begin{center}
		\includegraphics[width=0.405\textwidth]{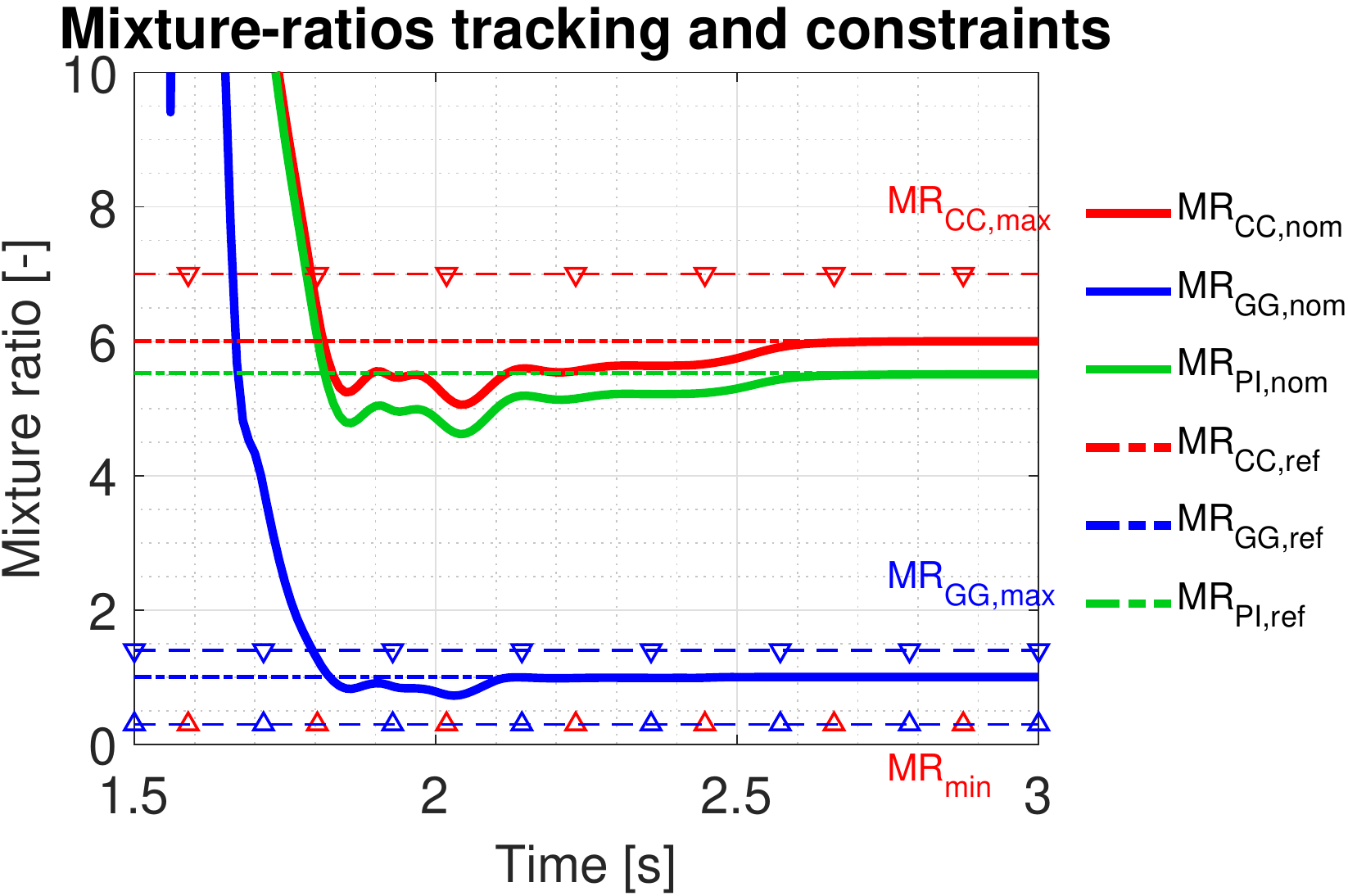}
		\caption{Tracking results in $MR$ for $p_{CC,r}=1$ (nominal)}
		\label{resu_MR}
	\end{center}
\end{figure} 
Mixture ratios naturally start from values very far from the allowable area, due to the low initial mass flows that hinder the definition of quotients. Indeed, chambers are not physically ignited during the first instants (even if igniters are active); hence, $MR$ are not relevant there. Fig. \ref{resu_pCC} depicts the results of $p_{CC}$ tracking for three operating points: $p_{CC,r}=1$ (nominal), $p_{CC,r}=0.7$ (minimum for this engine) and $p_{CC,r}=1.2$ (maximum). At all three points, the reference mixture ratios remain the same $MR_{CC,r}=6$, $MR_{GG,r}=1$ and $MR_{PI,r}=5.25$. $MR$ tracking for the nominal case is depicted in Fig. \ref{resu_MR}.\\
Tracking is achieved with sufficient accuracy in $p_{CC}$ for all cases (under $0.7$\%) and with little error in $MR$ (under $0.3$\% in nominal, under $1.7$\% in off-nominal) while respecting constraints up from the time when it is considered feasible and acceptable to respect them in practice ($1.9s$). 
The overshoot and oscillations present before achieving the final tracking are generated by the exogenous influence of the GG-starter input mass flow, which is not taken into account in the linearised model. Overshoot is more pronounced in the minimum case since the relative influence of the starter is more elevated.\\
The controller is able to achieve that tracking performance after random initial conditions coming from the sequential transient, whereas constraints-verification time oscillates some hundredths of seconds. Computational times in \textit{MATLAB} are of the order of ten times longer than real time, which does not rule out a future real-machine implementation.

\subsection{Comparison with open loop and other linear controllers}

\begin{figure}[t]
	\begin{center}
		\includegraphics[width=0.38\textwidth]{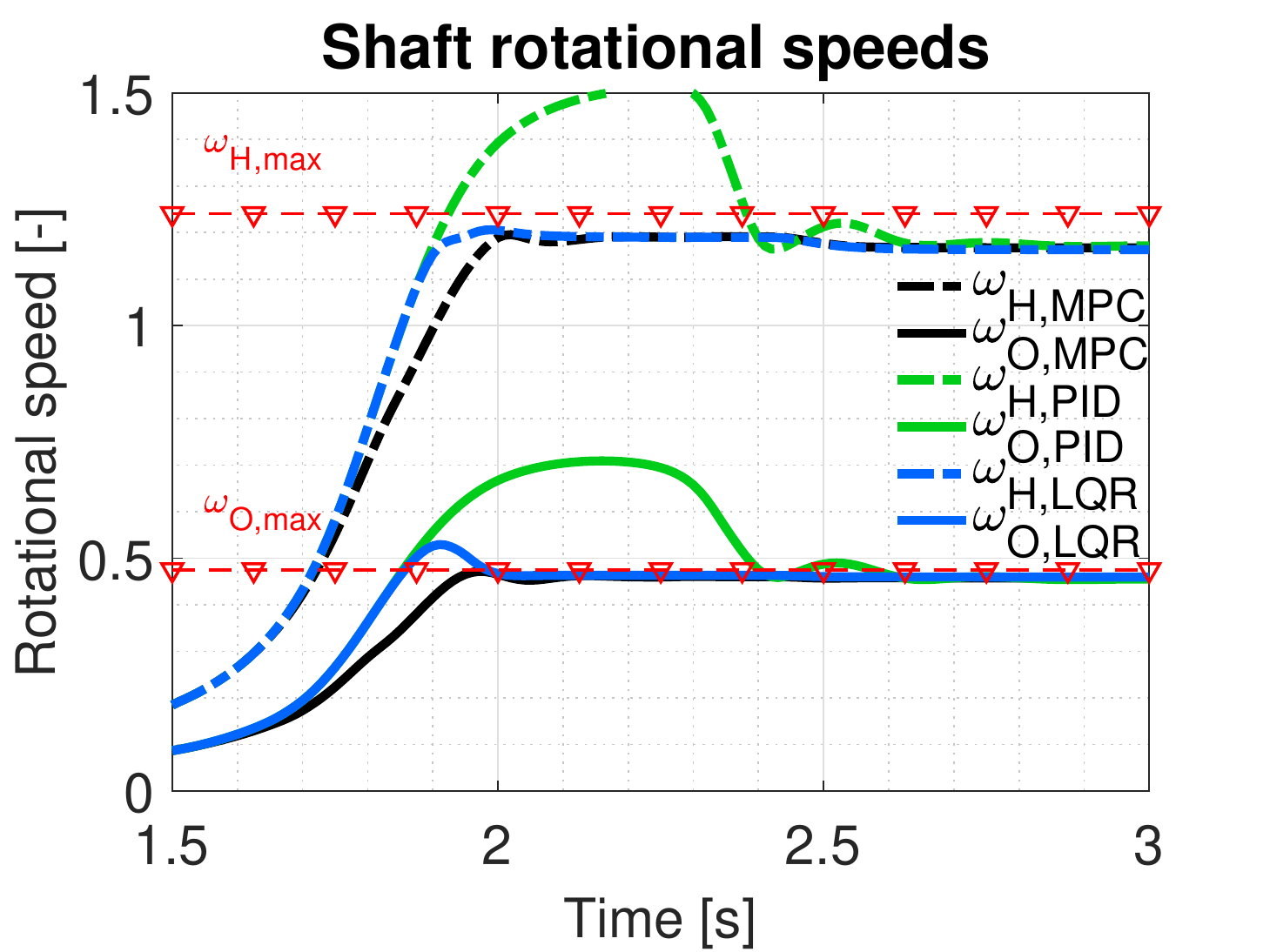}
		\caption{Rotational speeds  $\omega_H$ and $\omega_O$ for $p_{CC,r}=1.2$ with MPC, PID and LQR controllers}
		\label{resu_omegas}
	\end{center}
\end{figure}

Table \ref{tabOL} summarises the comparison between this closed-loop (CL) proposal and OL simulations in terms of some performance indicators.
\begin{table}[h]
	\caption{Performance-indicators comparison between this CL proposal and OL at the three selected operating points}
	\begin{center}
		\begin{tabular}{|p{3.4cm}|p{0.38cm}|p{0.38cm}|p{0.38cm}|p{0.48cm}|p{0.38cm}|p{0.49cm}|}
			\hline
			\textbf{Operating point} & \multicolumn{2}{|p{0.76cm}|}{\textbf{Nominal}} & \multicolumn{2}{|p{0.86cm}|}{\textbf{Minimum}} & \multicolumn{2}{|p{0.86cm}|}{\textbf{Maximum}} \\ \hline 
			\textbf{Indicator} & \textbf{OL}  & \textbf{CL} & \textbf{OL}  & \textbf{CL} & \textbf{OL}  & \textbf{CL}\\ \hline
			Settling time (99\%) $[s]$ & 2.8 & 2.51 & 2.67 & 2.55 & 2.69 & 2.53 \\ \hline
			Overshoot (\% in $p_{CC}$) & 6.31 & 5.04  & 15.1 & 11.46 & 3.34 & 4.04 \\ \hline
			Constraints verification $[s]$ & 1.81 & 1.8  & 1.83 & 1.76 & 1.77 & 1.81  \\
			\hline
			$p_{CC}$ static error (\%) & 0.25 & 0.26 & 2.8 & 0.26 & 0.34 & 0.67 \\ \hline
			$MR_{CC}$ static error (\%) & 0.17 & 0.01 & 2.58 & 1.38 & 3.18 & 1.37 \\ \hline
			$MR_{GG}$ static error (\%) & 1.39 & 0.05 & 1.31 & 0.69 & 1.23 & 0.59 \\ \hline
			$MR_{PI}$ static error (\%) & 1.43 & 0.3 & 2.84 & 0.85 & 3.41 & 1.64 \\
			\hline
		\end{tabular}
		\label{tabOL}
	\end{center}
\end{table}
The nominal OL is engine's original command, which is precisely tuned for the standard case. The minimum and maximum OL commands have been computed by means of the preprocessor explained in Section \ref{preproc}. The improvement with respect to OL is not dramatic, it is even worse in some indicators. Nonetheless, the real gain of this CL MPC control appears for operating points different from the nominal, where multivariable tracking was difficult to achieve with high performance while respecting constraints during the transient.\\ 
Moreover, other linear control methods have been tested on the same plant, such as simple PID and LQR controllers. Tracking results of these controllers are good in some of the reference variables (under $0.0001$\% in $p_{CC}$), but not for all of them simultaneously. Moreover, there are no guarantees of complying with all the constraints in this problem. Hence, when aiming at tracking off-nominal points, constraints are indeed highly violated. For instance, while throttling up until $p_{CC}=1.2$, the system controlled by PID or LQR has the tendency to surpass rotational speeds bounds, as depicted in Fig. \ref{resu_omegas}, whereas MPC respects them.
\section{Conclusion}\label{conclu}

The control of the transient phases of liquid-propellant rocket engines has traditionally been performed in open loop due to its highly nonlinear behaviour. This work has sought to improve the control of the fully continuous part of the start-up of a gas-generator-cycle LPRE, whose valves can be adjusted for controlling pressure in the main chamber and mass-flow mixture ratios. An MPC controller has been synthesised on that phase for tracking combustion-chamber pressure and mixture ratios while respecting a set of hard operational constraints. This controller is accompanied by a preprocessor that serves to provide a full-state reference built from launcher needs, by making use of a nonlinear state-space model of the engine. The linear MPC controller with integral action is able to track that end-state reference with sufficient accuracy and constraints are respected when necessary. Robustness, vital in this application with possible perturbations and internal-parameter variations, is taken into account for a given set of perturbation scenarios. The costly nested \textit{minimax} optimisation of typical robust MPC approaches has been rewritten as the minimisation of a scalar cost. In future work, other ways of posing this robustness consideration globally will be investigated. The tracking of a predefined trajectory will also be studied. A more extensive validation study with respect to perturbation cases will be carried out.


%
\end{document}